\documentclass[journal]{IEEEtran}

\ifCLASSINFOpdf
	\usepackage[pdftex]{graphicx}
    \graphicspath{{Figures/}}
    \DeclareGraphicsExtensions{.pdf}
\else
\fi
\usepackage{graphicx}
\ifCLASSOPTIONcompsoc
  \usepackage[caption=false,font=normalsize,labelfont=sf,textfont=sf]{subfig}
\else
  \usepackage[caption=false,font=footnotesize]{subfig}
\fi

\usepackage{amsfonts}
\usepackage{amsmath}
\makeatletter
\def\Let@{\def\\{\notag\math@cr}}
\makeatother
\usepackage{colortbl}
\usepackage{cite}
\usepackage{booktabs}
\usepackage{multirow}
\usepackage{algorithmic}
\usepackage{algorithm} 
\usepackage[shortlabels]{enumitem}
\usepackage{accents}
\usepackage{color}
\usepackage{amsmath}
\usepackage{amssymb}
\usepackage{amsthm,bm}

\usepackage{url}

\makeatletter
\def\bbordermatrix#1{\begingroup \m@th
  \global\let\perhaps@scriptstyle\scriptstyle
  \@tempdima 4.75\p@
  \setbox\z@\vbox{%
    \def\cr{%
      \crcr
      \noalign{%
        \kern2\p@
        \global\let\cr\endline
        \global\let\perhaps@scriptstyle\relax
      }%
    }%
    \ialign{$\make@scriptstyle{##}$\hfil\kern2\p@\kern\@tempdima
      &\thinspace\hfil$\perhaps@scriptstyle##$\hfil
      &&\quad\hfil$\perhaps@scriptstyle##$\hfil\crcr
      \omit\strut\hfil\crcr
      \noalign{\kern-\baselineskip}%
      #1\crcr\omit\strut\cr}}%
  \setbox\tw@\vbox{\unvcopy\z@\global\setbox\@ne\lastbox}%
  \setbox\tw@\hbox{\unhbox\@ne\unskip\global\setbox\@ne\lastbox}%
  \setbox\tw@\hbox{$\kern\wd\@ne\kern-\@tempdima\left[\kern-\wd\@ne
    \global\setbox\@ne\vbox{\box\@ne\kern2\p@}%
    \vcenter{\kern-\ht\@ne\unvbox\z@\kern-\baselineskip}\,\right]$}%
  \null\;\vbox{\kern\ht\@ne\box\tw@}\endgroup}
\def\make@scriptstyle#1{\vcenter{\hbox{$\scriptstyle#1$}}}
\makeatother

\newtheorem{theorem}{\textbf{Theorem}}

\newtheorem{lemma}{\textbf{Lemma}}
\renewenvironment{proof}{{\textbf{Proof of Lemma 1.}}}{\qed}

\begin{document}
	
\title{Leverage Point Identification Method\\for LAV-Based State Estimation}

\author{Mathias Dorier, Guglielmo~Frigo,~\IEEEmembership{Member,~IEEE,} Ali~Abur,~\IEEEmembership{Fellow Member,~IEEE} and Mario~Paolone,~\IEEEmembership{Senior Member,~IEEE}
}

\maketitle

\begin{abstract}
The state estimation problem can be solved through different methods. In terms of robustness, an effective approach is represented by the Least Absolute Value (LAV) estimator, though vulnerable to leverage points. Based on a previously proposed theorem, in this paper we enunciate, and rigorously demonstrate, a new lemma that proves the identifiability of leverage points in LAV-based state estimation. On the basis of these theoretical foundations, we propose an algorithm for leverage point identification whose performance is validated by means of extensive numerical simulations and compared against more traditional approaches, like Projection Statistics (PS). The obtained results confirm that the proposed method outperforms PS and represents a significant enhancement for LAV-based state estimators as it correctly identifies all the leverage points,
independently of the accuracy or the presence of measurement gross errors. A dedicated application example with respect to power system state estimation is finally included and discussed.
\end{abstract}


\IEEEpeerreviewmaketitle

\section{Introduction}
\label{sec:intro}
In power system state estimation a well-known challenge is represented by the correct identification of leverage points \cite{Abur-etAl1996, Mili-etAl1996, Zhao-etAl2018}. Indeed, even a single leverage point may affect the state estimation process with detrimental effects \cite{Abur1990, Majumdar-etAl2018}. 

Let us define the standard measurement model as a system of linear equations:
\begin{equation}
    \bm{H \bm{\theta + \varepsilon}} = \bm{z}
    \label{eq:intro_model}
\end{equation}
\begin{equation*}
    \bm{\theta} = \{\theta_k | k =1, \dots N \} \quad \bm{z}, \bm{\varepsilon} = \{z_i, \varepsilon_i | i =1, \dots M \}
\end{equation*}
where $\bm{z}$ and $\bm{\theta}$ are the vectors of the $M$ measurements and the $N$ unknown states, respectively. The matrix $\bm{H}$, of dimension $M \times N$, defines the relationship between measurements and states, whereas $\bm{\varepsilon}$ models the measurement uncertainty in terms of an additive random variable.

As known, the state estimation problem can be solved through different methods: Weighted Least Squares \cite{Abur-etAl2004}, standard and extended Kalman Filter \cite{Paolone-etAl2016,Ghahremani-etAl2011}, or Sparse State Recovery \cite{Xu-etAl2013}. In terms of robustness, though, the most effective approach is represented by the Least Absolute Value (LAV) estimator \cite{Abur-etAl1991} thanks to the automatic bad data rejection by proper measurement scaling \cite{Gol-etAl2014}. However, even the LAV estimator is vulnerable to leverage points.

In this regard, it is important to clarify the difference between outliers and leverage points. An outlier is a measurement $z_i$ that does not belong to the same statistical distribution of the other measurements $z_j$, with $j=1,...M, j\neq i$. Typically, an outlier is given by a gross error in the measurement acquisition process and produces a measurement that is inconsistent with the (known) statistical properties of the noise term $\bm{\varepsilon}$ of the measurement itself. A leverage point, instead, is a measurement $z_i$ that affects the state estimator in such a way that the state estimate will satisfy closely that particular measurement \cite{Rawlings-etAl1998}. It is also worth noticing that a leverage point is not necessarily associated with 
a wrong measurement, but if it coincides with a bad data, it affects significantly the estimation results \cite{Tan-etAl2014, Monticelli2004}.

Given the system $\bm{H}$ matrix, it is possible to derive the associated projection matrix of dimension $M\times M$ as:
\begin{equation}
    \bm{P = H (H^T H)^{-1} H^T}
    \label{eq:proj_matrix}
\end{equation}
whose diagonal terms $p_{ii}$ account for the influence of measurement $z_i$ on the state estimate $\hat{\bm{\theta}}$, and satisfy the following inequality \cite{Belsley-etAl1980}:
\begin{equation}
    0 \leq p_{ii} = diag(\bm{P}) \leq 1, \qquad ii = 1,\dots M
    \label{eq:infl_funct}
\end{equation}
The larger $p_{ii}$, the larger the influence. As a consequence, $p_{ii}$ values close to $1$ are typically associated with leverage points. Unfortunately, though, the identification of a leverage point based only on the projection matrix is not robust in the presence of a non-ideal and redundant $\bm{H}$ matrix \cite{Els-etAl1999}.


In LAV-based estimation, leverage points are associated with a more specific definition. Let us introduce the factor space $\bm{A}$ as the vector space spanned by matrix $\bm{H}$ rows $\bm{h}_i$:
\begin{equation}
    span(\bm{A}) = \left\{ \sum_{i=1}^{M} \alpha_i \bm{h}_i \Bigg| \alpha_i \in \mathbb{R}, \quad i = 1,\dots M\right\}
\end{equation}
In this context, a leverage point is defined as an outlier in the factor space. In the factor space, matrix $\bm{H}$ rows are expected to be located within a given restricted and well-defined area. As a consequence, also the corresponding measurements are characterized by a similar influence on the final state estimate $\hat{\bm{\theta}}$. However, if a row $\bm{h}_i$ proves to be an outlier in the factor space, i.e. it is located far from the cluster of the other $\bm{H}$ matrix rows or its subspace does not intersect with any other subspace in $\bm{A}$, the corresponding measurement $z_i$ will have a much higher influence on the final state estimate and has to be treated as a leverage point. As further discussed in the Appendix, the present definition is independent of the measurement accuracy, thus leverage points can be detected and identified whether they carry bad data or not.

The recent literature has proposed several methods for outliers' and leverage points' detection \cite{Rousseeuw1990}. Among the former ones, the Chi-Square ($\chi^2$) and the Largest Normalized Residuals (LNR) tests are the most widely employed \cite{Johnson-etAl2007, Aggarwal2017}. By assuming the residuals are normally distributed, the $\chi^2$ test determines a probability to decide whether a single measurement belongs to the same distribution as the other measurements. On the other hand, the LNR test proves to be effective in the presence of either single or multiple interacting but non-conforming\footnote{In this context, two bad data are said to be conforming if their residuals are consistent. For instance, in the power system state estimation scenario, if a set of bad data on power flows or power injections (nearly) satisfy the Kirchhoff power law, they are said to be conforming bad data \cite{Zhao-etAl2018}.} bad data \cite{Coutto-etAl2014,Asada-etAl2005}. Unfortunately, though, both $\chi^2$ and LNR tests fail in detecting leverage points due to their inherent reduced residual values, particularly if the measurements are characterized by low redundancy \cite{Zhao-etAl2018}.

Regarding the detection of leverage points, there exist several methods, most of them tailored to the adopted state model or estimator \cite{Rousseeuw-etAl1991, Mili-etAl1991, Abur-etAl1997}. In general, such methods attempt to quantify the influence of each measurement by means of statistical tests, e.g. projection statistics \cite{Mili-etAl1996_tsg}, or residual analysis, e.g. generalized Cook's distance \cite{Cook-etAl1980}.

In this paper, we consider the problem of leverage points' identification in LAV-based state estimation. 
In this context, we first introduce four main assumptions:
\begin{itemize}
    \item the model \eqref{eq:intro_model} is perfectly linear, i.e. the $\bm{H}$ matrix is independent of the state vector $\bm{\theta}$;
    \item the $\bm{H}$ matrix has full rank, i.e. all its columns are linearly independent;
    \item the measurement uncertainty $\bm{\varepsilon}$ consists of additive white Gaussian noise;
    \item the noise term is uncorrelated with the state vector and the time information.
\end{itemize}

Based on the theorem presented in \cite{Abur-etAl1996} (Theorem 1 at page 142), we enunciate and rigorously demonstrate a new lemma that proves the identifiability of leverage points in LAV-based state estimation. We also propose an algorithm that effectively implements the aforementioned theoretical results.

Furthermore, by means of extensive numerical simulations, we carry out a thorough characterization of the algorithm performance. The simulation results confirm how the proposed method significantly enhances the robustness of LAV-based estimators by guaranteeing full identifiability of leverage points.

The paper is organised as follows. In Section \ref{sec:literature}, we discuss the reference technique for leverage points' identification and its main applications. Section \ref{sec:method} introduces the theoretical foundations and describes in detail the proposed method. In Section \ref{sec:example} we validate its performance by means of numerical simulations inspired by state estimation applications in power systems. Section \ref{sec:concl} provides some closing remarks. Finally, in the Appendix we compare the performance of the proposed method against Projection Statistics in a simple yet significant application case.

\section{Traditional Approach for the\\Identification of Leverage Points}
\label{sec:literature}
In this Section, we briefly summarize the traditional approach towards leverage points' identification, namely projection statistics, and discuss its performance, advantages and still open issues.

Projection statistics has been introduced in \cite{Mili-etAl1996} as an alternative method to compute the distance in the factor space of each point (i.e. a single matrix $\bm{H}$ row) with respect to the cloud of the other points (i.e. all the other rows). In detail, each point is associated with the maximum value of the corresponding standardized projection, also referred to as \textit{projection statistic} (PS). The points and the corresponding PS values are expected to follow a Gaussian and a $\chi^2$ distribution, respectively,  but this does not apply to outliers \cite{Rousseeuw-etAl1991}. Therefore, $\chi^{2}_{d,0.975}$ is used as cutoff value, where $d$ is the number of degrees of freedom and $0.975$ is the quantile of the $\chi^2$ distribution. Any point exceeding this threshold is considered as an outlier, and thus identified as a leverage point. Thanks to its straightforward implementation, the PS method has been proven to be computationally efficient and compatible with real-time state estimation \cite{Els-etAl1999}.

The main advantage of the PS method consists of its reduced computational cost. On the other hand, a frequent drawback is the sparsity of the $\bm{H}$ matrix that might cause the differences between simple outliers and actual leverage points to be almost negligible. In this sense, the definition of the cutoff value typically relies on preliminary Monte-Carlo analysis that not only requires a prior knowledge of the system, but also risks to overfit the identification technique.

\section{Proposed Method}
\label{sec:method}
The Least Absolute Value (LAV) estimator determines the $N$ unknown states $\theta_k$ by minimizing the sum of the absolute values of the residuals $r_i$ of $M$ measurements $z_i$:
\begin{equation}
\hat{\bm{\theta}} = \mathrm{arg}\min_{\bm{\theta}} \sum_{i}^{M} |r_i| \qquad s.t. \quad \bm{z} = \bm{H} \cdot \hat{\bm{\theta}} + \bm{r}
\label{eq:method_eq1}
\end{equation}
where the residual $r_i$ depends on the measurement $z_i$ and the corresponding row $\bm{h}_i$ of the $\bm{H}$ matrix:
\begin{equation}
r_i = z_i - \bm{h}_i\cdot \hat{\bm{\theta}} = z_i - h_{i1}\cdot \hat{\theta}_1 - \dots - h_{iN}\cdot \hat{\theta}_N
\label{eq:method_eq2}
\end{equation}

Then, the objective function in \eqref{eq:method_eq1} can be written as follows:
\begin{equation}
		\sum_{i}^{M} |r_i| = \sum_{i}^{M} |z_i - \bm{h}_i\cdot \hat{\bm{\theta}}| = \sum_{i}^{M} f_i(\hat{\bm{\theta}})
	\label{eq:method_eq3}
\end{equation}

Based on Jensen's inequality, $f_i$ functions prove to be convex. Given a weight $t\in[0, 1]$ and two state estimates $\hat{\bm{\theta}}_1$ and $\hat{\bm{\theta}}_2$:
\begin{eqnarray}
		f_i\left(t\cdot \hat{\bm{\theta}}_1 + (1-t)\cdot \hat{\bm{\theta}}_2\right) \underset{t=\frac{1}{2}}{=} \frac{1}{2} |2\cdot z_i - \bm{h}_i\cdot \hat{\bm{\theta}}_1- \bm{h}_i\cdot \hat{\bm{\theta}}_2|  \nonumber \\
		\leq \frac{1}{2} \cdot \left(|z_i - \bm{h}_i\cdot \hat{\bm{\theta}}_1| + |z_i - \bm{h}_i\cdot \hat{\bm{\theta}}_2|\right) \\
		\Rightarrow f_i\left(\frac{1}{2}\cdot \hat{\bm{\theta}}_1 + \frac{1}{2}\cdot \hat{\bm{\theta}}_2\right) \leq \frac{1}{2}\cdot f_i(\hat{\bm{\theta}}_1) + \frac{1}{2}\cdot f_i(\hat{\bm{\theta}}_2) \nonumber
	\label{eq:method_eq35}
\end{eqnarray}
and a similar result holds for the sum $F$ of the $M$ $f_i$ functions:
\begin{eqnarray}
			F\left(\frac{1}{2}\cdot \hat{\bm{\theta}}_1 + \frac{1}{2}\cdot \hat{\bm{\theta}}_2\right) = \sum_{i}^{M} f_i\left(\frac{1}{2}\cdot \hat{\bm{\theta}}_1 + \frac{1}{2}\cdot \hat{\bm{\theta}}_2\right) \nonumber \\ 
			\leq \sum_{i}^{M} \left(\frac{1}{2}\cdot f_i(\hat{\bm{\theta}}_1) + \frac{1}{2}\cdot f_i(\hat{\bm{\theta}}_1)\right) \\
			= \frac{1}{2}\cdot \left[ \sum_{i}^{M} f_i(\hat{\bm{\theta}}_1) + \sum_{i}^{M} f_i(\hat{\bm{\theta}}_2)\right] = \frac{1}{2} \cdot \left(F(\hat{\bm{\theta}}_1)+ F(\hat{\bm{\theta}}_2) \right)\nonumber
	\label{eq:method_eq375}
\end{eqnarray}

Based on \eqref{eq:method_eq35} and  \eqref{eq:method_eq375}, the objective function of the LAV estimation problem \eqref{eq:method_eq1} proves to be convex. This fundamental property will be used in Section III.A.

It is worth noticing that $F$ is convex, but not strictly convex. In fact, there might exist particular combinations of $f_i$ functions that cause the slope of the sum function to be zero. In this case, there might be multiple minima and the identification process will end in one of these points depending on the initial conditions of the optimization problem. This specific condition occurs mostly 
when there are not enough measurements. Accordingly, it is sufficient to add further measurements and thus 
guarantee the strict convexity. Nevertheless, as shown in the Appendix, the occurrence of such a condition does not limit the applicability of the proposed method. It means only that one or more $\bm{H}$ matrix rows lie on the boundary between being considered leverage points and the factor space cluster. To identify them as leverage points represents a conservative approach and guarantees a more accurate and reliable state estimation.

It is also interesting to observe that the zero of $f_i(\hat{\bm{\theta}})$ is a locus with $N-1$ degrees of freedom. Therefore, any state $\hat{\theta}_k$ can be expressed as a linear function of the other states:
\begin{eqnarray}
	f_i(\hat{\bm{\theta}}) = 0 \nonumber \\
	\Leftrightarrow z_i - h_{i1}\cdot \hat{\theta}_1 - \hdots - h_{iN}\cdot \hat{\theta}_N = 0\\
	\Leftrightarrow \hat{\theta}_k = \frac{z_i - h_{ij}\cdot \hat{\theta}_j}{h_{ik}}, \quad j \in [1, \dots N] \wedge j \neq k \nonumber 
\label{eq:method_eq4}	
\end{eqnarray}

As a consequence, given $M \geq N$ measurements, all the intersections between the functions $f_i$ are defined by at least $N$ of these loci, and are zeros of these functions. Nevertheless, since the objective function is the sum of the functions $f_i$, the minimum of the sum must coincide with one of these zeros, and must be the global minimum \cite{Barrodale-etAl1973}. 

\subsection{Theoretical Foundations}
\noindent \textbf{Hypothesis.} \textit{Given the following three assumptions:
\begin{itemize}
\item The state estimate $\hat{\bm{\theta}}$ is given by a LAV estimator.
\item The $\bm{H}$ matrix is known and refers to a linear or linearized system.
\item The $\bm{H}$ matrix does not vary during the application of the method.
\end{itemize}}

\begin{theorem}
\textit{If the $\bm{H}$ matrix column rank  is equal to $L \leq N$, then there exists a LAV estimate which satisfies at least $L$ of the measurements $z_i$ exactly (with zero residuals)} \cite{Abur-etAl1996}\footnote{The proof of Theorem 1 is given in \cite{Abur-etAl1996} and constitutes also the basis for the proof of the following Lemma 1.}.
\end{theorem}

As a consequence, if the $\bm{H}$ matrix is full-rank, there are at least $N$ measurements among $M$ that correspond to the true state values and produce a zero residual. The following lemma can be thus introduced:
\begin{lemma}
\textit{Let $\bm{H}$ be the matrix that describes the exact relationship between the $N$ states and $M$ measurements of a linear system, and $\bm{h}_i$ the $i^{th}$ row of the $\bm{H}$ matrix. If a row $\bm{h}_j$ of the $H$ matrix is a leverage point, then there exists a linear combination of $N-1$ other rows that satisfies the following inequality:}
\begin{equation}
		s = \sum_{i \neq j}^{M} |\bm{h}_i \cdot \bm{v}| \leq |\bm{h}_j \cdot \bm{v}| = q
	\label{eq:method_lemma}
\end{equation}
\textit{where $\bm{v}$ is the unitary vector that completes the basis $\bm{B}$ formed by all selected $N-1$ rows.}
\end{lemma}

Given $N$ measurements that satisfy Theorem 1, Lemma 1 guarantees that, if a measurement is a leverage point, it can be correctly identified based on the other $N-1$ measurements.

\noindent\begin{proof}
Let us consider the case where all measurements $z_i$ satisfy perfectly a set of states $\bm{\theta}^{true}$, except one measurement $z_j$ corresponding to row $\bm{h}_j$ that satisfies the set of states $\bm{\theta}^{err}$, as a result of a measurement gross error. 

After running the LAV estimator, if the estimated states $\hat{\bm{\theta}}$ are equal to $\bm{\theta}^{true}$, the residuals $r_i$ will be equal to zero, except for the point $\bm{h_j}$. Conversely, if the estimated states $\hat{\bm{\theta}}$ are equal to $\bm{\theta}^{err}$, only the residual $r_j$ will be equal to zero, and $\bm{h}_j$ has to be identified as a leverage point.

If it is possible to reject the gross error and restore the correct state estimation $\hat{\bm{\theta}} = \bm{\theta}^{true}$, the residuals' sum yields:
\begin{eqnarray}
	\sum_{i}^{M} |r_i| &=& \sum_{i}^{M} |\bm{h}_i \cdot \hat{\bm{\theta}} - z_i| \nonumber \\ 
	&=& \sum_{i \neq j}^{M} |\bm{h}_i \cdot \hat{\bm{\theta}} - z_i| + |\bm{h}_j \cdot \hat{\bm{\theta}} - z_j| \\ 
	&=& |\bm{h}_j \cdot \bm{\theta}^{true} -\bm{h}_j \cdot \bm{\theta}^{err}| = |\bm{h_j} \cdot \bm{\Delta \theta}| \nonumber
	\label{eq:method_proof1}
\end{eqnarray}

If the gross error is not properly neutralized, instead, all the $f_i(\hat{\bm{\theta}})$ zeros will intersect at the same point, except for the one associated to the leverage point, namely $f_j(\hat{\bm{\theta}})$. In other words, the minimization problem  has converged to the global minimum that contains inevitably the locus of the leverage point. 

Indeed, if $\hat{\bm{\theta}}=\bm{\theta}^{err}$, the residuals' sum is given by:
\begin{eqnarray}
		\sum_{i}^{M} |r_i| &=& \sum_{i}^{M} |\bm{h}_i \cdot \hat{\bm{\theta}}- z_i| \nonumber\\ 
		&=& \sum_{i \neq j}^{M} |\bm{h}_i \cdot \hat{\bm{\theta}}- z_i| + |\bm{h}_j \cdot \hat{\bm{\theta}}-z_j| \\ 
		&=& \sum_{i \neq j}^{M} |\bm{h}_i \cdot \bm{\theta}^{err} -\bm{h}_i \cdot \bm{\theta}^{true}| = \sum_{i \neq j}^{M} |\bm{h}_i \cdot \bm{\Delta \theta}| \nonumber
	\label{eq:method_proof2}
\end{eqnarray}

In this regard, it should be noticed that the LAV estimator minimizes the residuals' sum (in accordance with \eqref{eq:method_eq1}), but does not guarantee the uniqueness of the obtained minimum\footnote{As aforementioned, based on (9) and (10), the objective function in (5) is proved to be convex, but not strictly convex. Nevertheless, the unlikely condition of multiple minimum points is associated to the presence of leverage points on the boundary of the subspace spanned by the other matrix $\bm{H}$ rows and has to be suitably addressed, as shown by the example in the Appendix.}, as shown by the following inequality:
\begin{equation}
	\sum_{i \neq j}^{M} |\bm{h}_i \cdot \bm{\Delta \theta}| \leq |\bm{h}_j \cdot \bm{\Delta \theta}|
	\label{eq:method_proof3}
\end{equation}

According to Theorem 1, at least $N$ residual terms are equal to zero: one is associated with the leverage point, namely $r_j$; and the others correspond to $N-1$ rows $\bm{h'}_i$ such that:
\begin{equation}
	r_i' = \bm{h'}_i \cdot \bm{\Delta \theta} = 0 \Leftrightarrow \bm{h'}_i~ \bot~ \bm{\Delta \theta} 
	\label{eq:method_proof4}
\end{equation}

In other words, there exist at least $N-1$ rows $\bm{h'}_i$ perpendicular to the state error $\bm{\Delta \theta}$, that is defined in a $N$ dimensional vector space. Hence, $\bm{\Delta \theta}$ is also collinear with the unitary vector $\bm{v}$ that completes the basis $\bm{B}$ formed by all $N-1$ vectors $\bm{h'}_i$:
\begin{equation}
		\bm{v} = null(\bm{h'}_1, \hdots \bm{h'}_i, \hdots \bm{h'}_{N-1}), \quad \|\bm{v}\|= 1
	\label{eq:method_proof5}
\end{equation}
Consequently, the state error $\bm{\Delta \theta}$ can be also expressed as:
\begin{equation}
		\bm{\Delta \theta} = \bm{v}\cdot \|\bm{\Delta \theta}\|
	\label{eq:method_proof6}
\end{equation}

By substituting \eqref{eq:method_proof5} and \eqref{eq:method_proof6} in the inequality \eqref{eq:method_proof3}, we get:
\begin{eqnarray}
		\sum_{i \neq j}^{M} |\bm{h}_i \cdot \bm{\Delta \theta}| &\leq& |\bm{h}_j \cdot \bm{\Delta \theta}| \nonumber \\
		\sum_{i \neq j}^{M} |\bm{h}_i \cdot \bm{v}|\cdot ||\bm{\Delta \theta}|| &\leq& |\bm{h}_j \cdot \bm{v}|\cdot ||\bm{\Delta \theta}|| \\
		\sum_{i \neq j}^{M} |\bm{h}_i \cdot \bm{v}| &\leq& |\bm{h}_j \cdot \bm{v}| \nonumber
	\label{eq:method_proof7}
\end{eqnarray}
\end{proof}{}

It is worth noticing that the proved Lemma guarantees full-identifiability of the any leverage point, independently of the actual state estimates provided by the LAV estimator. In this sense, its application does not introduce any constraints on the estimator accuracy or on the number of leverage points.

\subsection{Method Implementation}
Based on Lemma 1, we developed an algorithm for the identification of leverage points, whose main processing stages are summarized in Algorithm 1. In particular, given the system $\bm{H}$ matrix, the algorithm identifies the set of rows $\bm{h}_j$ that correspond to leverage points and bias the state estimation.

First, we select a set of $N-1$ rows $\bm{h}_i$, that constitute a vector basis $\bm{B}$ (line 1). Then, we complete the basis by including a unitary vector $\bm{v}$ that is orthogonal to all the selected rows (line 2). Given a generic row index $j$, we compute the linear combination $s$ (line 3). If the inequality \eqref{eq:method_lemma} is verified, the corresponding row $\bm{h}_j$ is identified as a leverage point, otherwise we proceed to the next iteration (line 4). Once all the possible combinations are processed, all the identified leverage points are collected together in the set $S$ (line 5). This information can be employed to reject the corresponding measurements $z_j$ and get a refined state estimate.
\begin{algorithm}
		\caption{Leverage Point Identification}
		\label{alg:method}
		\begin{algorithmic}
			\STATE \textbf{Input:} $ \bm{H} $
			\STATE \textbf{Output:} $ S $
			\STATE 1. subspace with $N-1$ dof: $\bm{B}$ = $\bigcup$ $\bm{h}_i$, $i=1, \dots N-1$
			\STATE 2. normalized perpendicular vector: $\bm{v} \perp \bm{B}, ||v|| = 1$ 
			\STATE \hspace{0.25cm} \textbf{for} $j = 1, \dots M$
			\STATE 3. projection sum: $s$ = $\sum$ $|\bm{h}_i \cdot \bm{v}|$, $i \neq j$
			\STATE \hspace{0.75cm} \textbf{if} $s < |\bm{h}_j \cdot \bm{v}|$
			\STATE 4. leverage point identification: $\bm{h}_j$ $\rightarrow$ $L$
			\STATE \hspace{0.75cm} \textbf{end}	
			\STATE \hspace{0.25cm} \textbf{end}
			\STATE 5. identified leverage points: $\forall \bm{h}_j \in S$
		\end{algorithmic}
	\end{algorithm}

\section{Application Example}
\label{sec:example}
\begin{figure}
	\centering
		\includegraphics[width=.95\columnwidth]{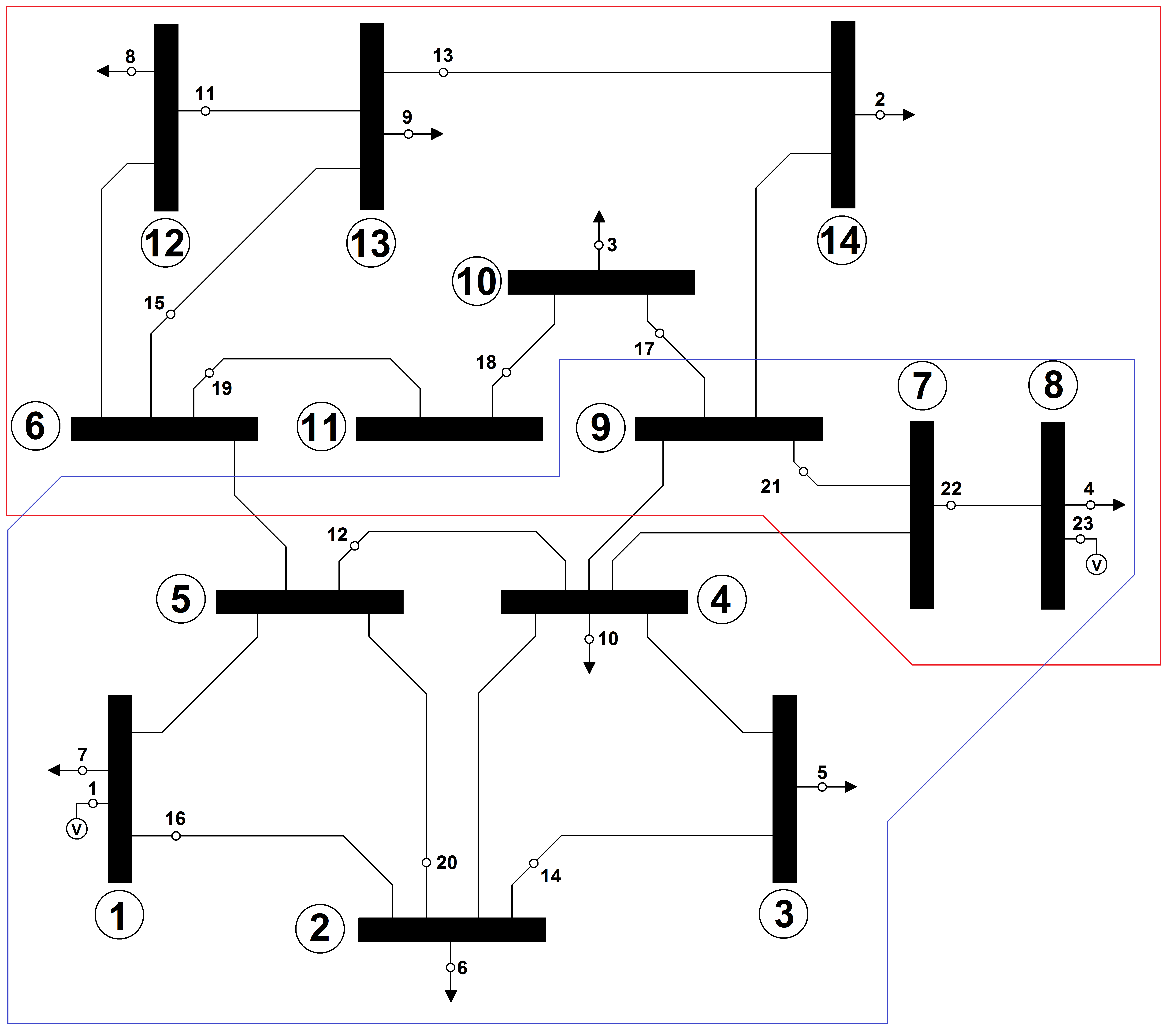}
		\caption{One-line diagram of the IEEE 14 Bus Test Case \cite{Ieee14bus}.}
		\label{fig:14BusSystem}
\end{figure}
In order to validate the accuracy and reliability of the proposed method, we carried out extensive numerical simulations on the IEEE 14 Bus Test Case \cite{Ieee14bus}. In this case, the measurement model is derived from the linearized load flow equations and thus fully satisfies hypothesis in Section III.A. As shown in Fig. \ref{fig:14BusSystem}, this benchmark system represents a portion of the American Electric Power System (in the Midwestern US) as of February, 1962. 
It consists of 14 buses, 5 generators, and 11 loads, and is equipped with 42 active and reactive power and 2 voltage measurement points, i.e. $M = 44$. The system state vector of dimension $N \times 1$ contains the bus voltage magnitudes and phase angles for each bus where $N=28$. 

If bus $1$ is taken as the angle reference, the number of unitary vectors to compute per measurement is equal to:
\begin{eqnarray}
    Card\left(\bigcup~ \bm{v}\right) = \frac{(M-1)!}{((M-1)-(N-1))! \cdot (N-1)!} \nonumber \\ 
	 = \frac{(43-1)!}{((43-1)-(27-1))! \cdot (27-1)!} = 166.51e9
\end{eqnarray}
This number of iterations is too high and would take a prohibitively long time to be computed. In order to overcome this issue, the system is divided into multiple subsystems. The basic idea is that 
two separate subsystems that don't have any states in common will not intersect in the state space, and thus, will have no influence on each other. 

In Fig. \ref{fig:14BusSystem} we delimit the two subsystems within red and blue contours, respectively. In order to reproduce a plausible operating condition, gross errors have been added to the measurements. In Table \ref{table:SystemResults}, the first column defines which measurements act as leverage points, whereas the second and third columns report the identification results when running the proposed method separately over the blue and red system partitioning, respectively.
\begin{table}
    \caption{IEEE 14 Bus Test Case - Leverage Point Identification}
	\label{table:SystemResults}
	\centering
		\begin{tabular}{c||c|c|c}
			\toprule
			\textbf{Meas.} & \textbf{Add. GE} & \textbf{LPs Syst. blue} & \textbf{LPs Syst. red}\\ \midrule
			$|V_1|$	& Unbiased Est. & -& $\varnothing$ \\ 
			$P_{inj3}$ & \textbf{Biased Est.} & Leverage Point & $\varnothing$ \\ 
			$Q_{inj3}$ & \textbf{Biased Est.} & Leverage Point& $\varnothing$ \\ 
			$P_{inj2}$ & \textbf{Biased Est.} & Leverage Point& $\varnothing$ \\ 
			$Q_{inj2}$ & \textbf{Biased Est.} & Leverage Point& $\varnothing$ \\ 
			$P_{inj1}$ & Unbiased Est. & -& $\varnothing$ \\ 
			$Q_{inj1}$ & Unbiased Est. & -& $\varnothing$ \\ 
			$P_{inj4}$ & \textbf{Biased Est.} & Leverage Point& $\varnothing$  \\ 
			$Q_{inj4}$ & \textbf{Biased Est.} & Leverage Point& $\varnothing$  \\ 
			$P_{flow5-4}$ & \textbf{Biased Est.} &Leverage Point& $\varnothing$ \\ 
			$Q_{flow5-4}$ & \textbf{Biased Est.} & Leverage Point& $\varnothing$ \\ 
			$P_{flow2-3}$ & Unbiased Est. & -& $\varnothing$ \\ 
			$Q_{flow2-3}$ & Unbiased Est. & -& $\varnothing$ \\ 
			$P_{flow1-2}$ & Unbiased Est. & -& $\varnothing$ \\ 
			$Q_{flow1-2}$ & Unbiased Est. & -& $\varnothing$ \\ 
			$P_{flow2-5}$ & Unbiased Est. & -& $\varnothing$ \\ 
			$Q_{flow2-5}$ & Unbiased Est. & -& $\varnothing$ \\ 
			$P_{inj14}$ & \textbf{Biased Est.} & $\varnothing$& Leverage Point  \\ 
			$Q_{inj14}$ & \textbf{Biased Est.} & $\varnothing$&Leverage Point   \\ 
			$P_{inj10}$ & Unbiased Est. & $\varnothing$& -  \\ 
			$Q_{inj10}$ & Unbiased Est. & $\varnothing$& -  \\ 
			$P_{inj8}$ & \textbf{Biased Est.} & Leverage Point&Leverage Point  \\ 
			$Q_{inj8}$ & Unbiased Est. & - & - \\ 
			$P_{inj12}$ & \textbf{Biased Est.} & $\varnothing$&Leverage Point  \\ 
			$Q_{inj12}$ & \textbf{Biased Est.} & $\varnothing$&Leverage Point \\ 
			$P_{inj13}$ & Unbiased Est. & $\varnothing$& -  \\ 
			$Q_{inj13}$ & Unbiased Est. & $\varnothing$&  -\\ 
			$P_{flow12-13}$ & Unbiased Est. & $\varnothing$& - \\ 
			$Q_{flow12-13}$ & Unbiased Est. & $\varnothing$& - \\ 
			$P_{flow13-14}$ & Unbiased Est. & $\varnothing$& -  \\ 
			$Q_{flow13-14}$ & Unbiased Est. & $\varnothing$& -  \\ 
			$P_{flow6-13}$ & Unbiased Est. & $\varnothing$& - \\ 
			$Q_{flow6-13}$ & Unbiased Est. & $\varnothing$& - \\ 
			$P_{flow10-9}$ & Unbiased Est. & $\varnothing$& - \\ 
			$Q_{flow10-9}$ & Unbiased Est. & $\varnothing$& - \\ 
			$P_{flow11-10}$ & Unbiased Est. & $\varnothing$& - \\ 
			$Q_{flow11-10}$ & Unbiased Est. & $\varnothing$& - \\ 
			$P_{flow6-11}$ & \textbf{Biased Est.} & $\varnothing$&Leverage Point \\ 
			$Q_{flow6-11}$ & \textbf{Biased Est.} & $\varnothing$&Leverage Point \\ 
			$P_{flow9-7}$ & \textbf{Biased Est.} & Leverage Point&Leverage Point \\ 
			$Q_{flow9-7}$ & \textbf{Biased Est.} &Leverage Point & Leverage Point\\ 
			$P_{flow7-8}$ & Unbiased Est. &Leverage Point  & Leverage Point\\ 
			$Q_{flow7-8}$ & Unbiased Est. & - & -\\ 
			$|V_8|$ & Unbiased Est. & - &Leverage Point  \\ \bottomrule
		\end{tabular}
	
\end{table}

As evident from these results, the proposed method successfully identifies all the leverage points by using the system partitioning. A peculiar case is represented by $P_{flow7-8}$ that is classified as a leverage point despite not being affected by a gross error. Even if this might seem a deficiency of the identification technique, in practice there exists a clear motivation for such result. In the factor space, the measurement $P_{flow7-8}$ lies on the boundary between being a leverage point and belonging to the cloud of the other $\bm{H}$ matrix rows. Its physical counterpart $P_{inj8}$ is corrupted by a gross error, whereas $P_{flow7-8}$ itself is error free. In this limit condition, the convergence of the LAV estimator is susceptible to its initialization and even measurements that are not strictly leverage points might produce a significant deviation of the final state estimate. For this reason, it is reasonable to classify $P_{flow7-8}$ as a leverage point and neutralize its effect on the LAV estimator (as a sort of precautionary action).

It is also worth pointing out that not any partitioning guarantees the full identifiability. Let us consider the voltage magnitude measurements $|V_8|$ and $|V_1|$. If they both belong to the same subsystem, they will be correctly classified as unaffected measurements. However, if they are separated into different subsystems, the method might fail and identify one of the two as a leverage point. This case suggests that the analysis should be repeated with different system partitioning and the results must be compared in case of inconsistent measurement classifications.

Based on these observations, the proposed method proves to be an effective and reliable solution for the identification of leverage points. In particular, all the leverage points have been correctly identified and the two false positives can be explained by boundary effects or  inefficient system partitioning. 

\section{Conclusions}
\label{sec:concl}
The correct identification of leverage points represents a still unresolved challenge in LAV-based state estimation. The recent literature has proposed several methods that attempt to quantify the influence associated to each measurement by means of statistical tests, as for Projection Statistics, or residual analysis as for Generalized Cook's Distance. However, most of the proposed solutions are tailored to the specific system under analysis and requires a preliminary analysis for the definition of the cutoff value. 

Based on a previously proposed theorem, in this paper, we present and rigorously demonstrate a new lemma for the full identifiability of leverage points in LAV-based state estimation. Based on these theoretical foundation, we develop an effective identification method that does not introduce any constraint on the estimator accuracy or on the number of leverage points. 

The proposed method is thoroughly characterized by means of numerical simulations inspired by state estimation applications in power systems. In this context, we discuss also the unlikely conditions associated to leverage points lying in a boundary region or inefficient system partitioning. Based on the obtained results, it is reasonable to say that the proposed method proves to be an effective and reliable solutions as it correctly identifies all the leverage points, based only on the knowledge of the $\bm{H}$ matrix.

In the Appendix, we compare  the  performance  the  proposed method against Projection Statistics in a simple yet significant application case. Differently from the traditional approach, the proposed method allows to correctly classify all the $\bm{H}$ matrix rows, independently from the presence of measurement gross errors. For this analysis, we consider also the case of PMU-based measurements, where the $\bm{H}$ matrix is truly linear and is not a linearized approximation of the relationship between states and measurements.

\section*{Appendix: Comparison with PS}

\begin{figure}
\centering
	  \includegraphics[width=.7\columnwidth]{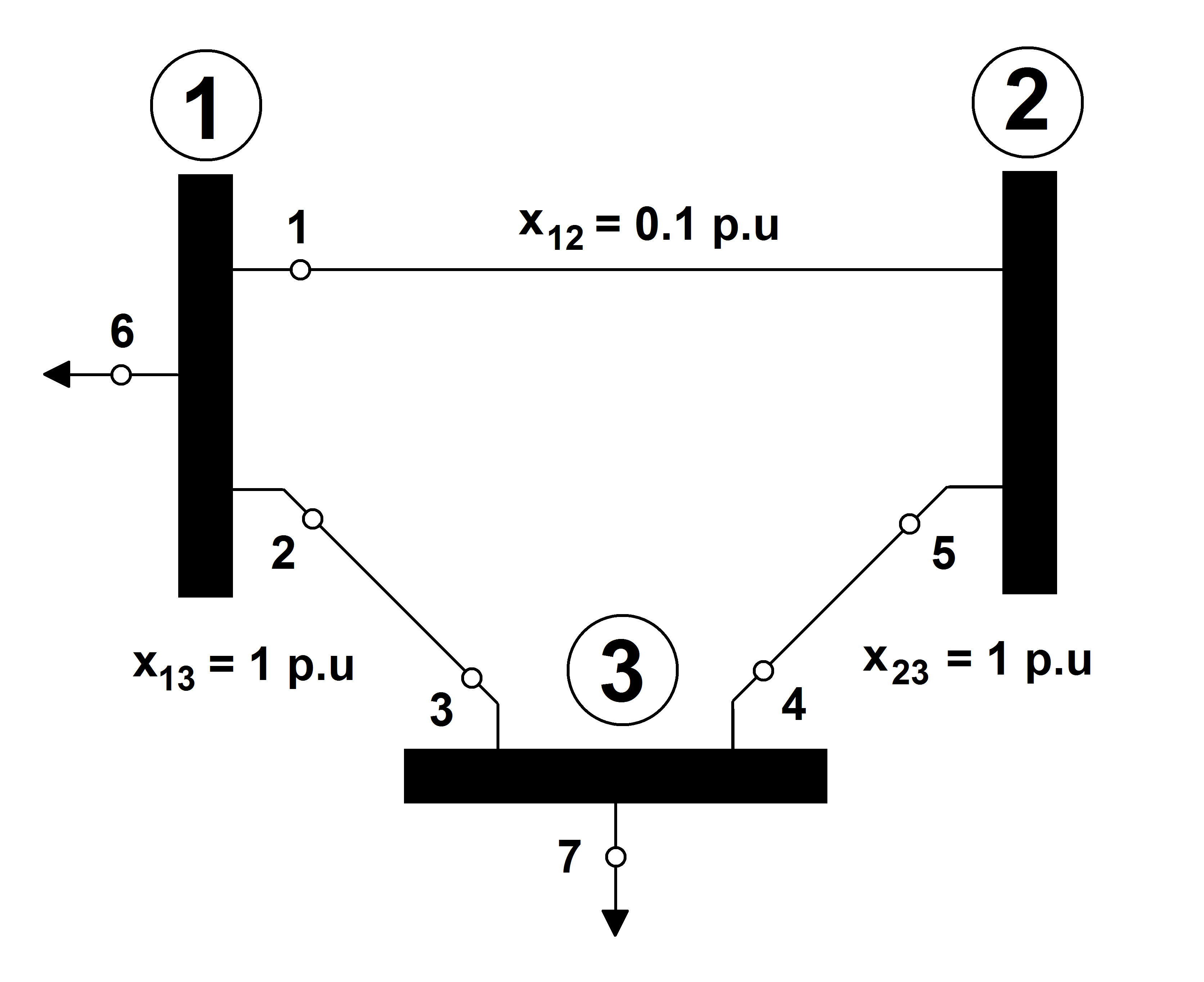}
		\caption{Schematic of the 3-Bus system presented in \cite{Mili-etAl1996}.}
		\label{fig:3BusSystem}
\end{figure}

In Fig. \ref{fig:3BusSystem}, we present the 3 Bus system, originally introduced in \cite{Mili-etAl1996}, that represents an ideal test bench to compare the proposed method against the traditional PS approach. In this case, the system states are the phase angles $\varphi_i$ for the $i^{th}$ node, the node voltages are assumed to have unitary amplitude. The lines are assumed to have a null resistance, and a pure longitudinal reactance. The measurements consist of 5 power flows and 2 power injections. It is then possible to compute the system $\bm{H}$ matrix using the DC-flow approximation:
\begin{equation}
    \bm{H} = \begin{bmatrix}
            10 & 1 & -1 & 0 & 0 & 11 & -1\\
            -10 & 0 & 0 & -1 & 1 & -10 & -1\\
    \end{bmatrix}^{T}
\end{equation}
where the phase $\varphi_3$ is taken as the reference. 

In Table \ref{TablePS1}, we report the PS results in terms of projection statistic, $\chi^{2}$ test and the corresponding degrees of freedom. 
\begin{table}
	\centering
	\caption{Projection Statistics Results}
	\label{TablePS1}
	\begin{tabular}{c|c|c|c}
		\toprule
		Meas. \# & PS     & $\chi^{2}_{d,0.975}$  & $d$ \\
		\midrule
		1        & 16.77  & 7.378 & 2  \\ 
		2        & 0.839  & 5.024 & 1  \\ 
		3        & 0.839  & 5.024 & 1   \\ 
		4        & 0.839  & 5.024 & 1   \\ 
		5        & 0.839  & 5.024 & 1   \\ 
		6        & 17.609 & 7.378 & 2  \\ 
		7        & 1.677  & 7.378 & 2  \\ 
		\bottomrule
	\end{tabular}
	
\end{table}
These results show that PS identifies as leverage points the $\bm{H}$ matrix rows associated to $P_{flow 1,2}$ and $P_{inj1}$. As a consequence, it is reasonable to expect that a gross error in these measurements would deviate the estimates from the true states. However, the LAV-based estimates do not deviate from the true states, independently from the gross error level. Therefore, it is reasonable to say that these $\bm{H}$ matrix rows should not be identified as leverage points.

\begin{figure}
	\begin{center}
		\includegraphics[width=.7\columnwidth]{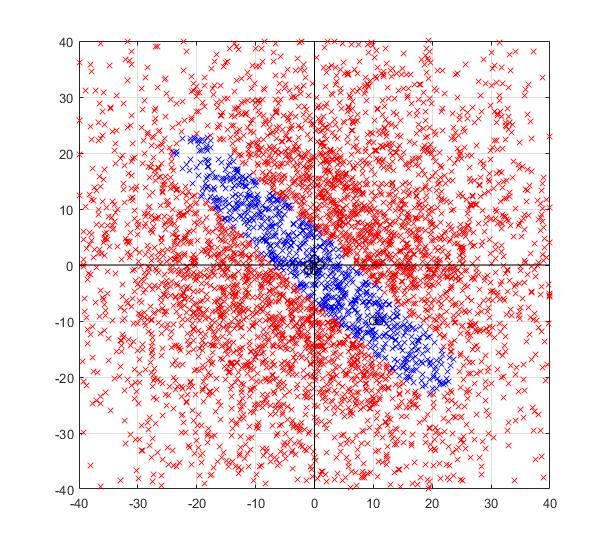}
		\caption{Factor for successful (in blue) and unsuccessful (in red) estimations with a gross error of 10 p.u. The $x$ and $y$ axes refer to $h_i1$ and $h_i2$, respectively.}
		\label{3BusSystemRandomMeasurement}
	\end{center}
\end{figure}

In order to understand what makes a measurement a leverage point, the following test has been conducted: an additional row $\bm{h}_8$ is added to the $\bm{H}$ matrix of the initial 3 bus system, whose parameters follow a normal distribution with mean $\mu = 0$ and variance $\sigma^2 = 30$. The true states $\bm{\theta}^{true}$ follow a normal distribution with $\mu = 0$ and $\sigma^2 = 1$, and the measurements are computed based on these states. A gross error of 10 p.u. is then added to the measurement of this random row.  Then, the LAV estimator is ran multiple times for random values of the last row and states. In Fig. \ref{3BusSystemRandomMeasurement}, we plot the initial rows of the $\bm{H}$ matrix in the factor space, as well as all the additional random lines. If the estimation is successful, the random lines are plotted in blue, otherwise in red.
    
There is a clear geometrical limit between the rows of $\bm{H}$ that deviate the estimation or not. These limits are based on the nature of the LAV estimator, and they are described according to Lemma 1. In Table \ref{TablePS2}, we report the two sides $s$ and $q$ of the inequality \eqref{eq:method_lemma} in Lemma 1 for all the rows and 5 different realizations of $\bm{v}$.
The results show that, in the LAV estimation problem, there are no leverage points detected, which matches the previous observations.

\begin{table*}
	\centering
	\caption{Proposed Method applied to the $\bm{H}$ matrix of the 3 bus system}
	\label{TablePS2}
	\begin{tabular}{c||c|c||c|c||c|c||c|c||c|c}
	\toprule
		        \multicolumn{1}{c||}{} & \multicolumn{2}{c||}{(0.707;0.707)} & \multicolumn{2}{c||}{(0,1)} & \multicolumn{2}{c||}{(1,0)} & \multicolumn{2}{c||}{(0.673;0.74)} & \multicolumn{2}{c}{(-0.707;0.707)} \\
		        & $s$ & $q$ & $s$ & $q$ &  $s$ & $q$ &  $s$ & $q$ &  $s$ & $q$ \\ 
		        \midrule
		$\bm{h_1}$ & 0 & 4.95 & 10 & 13 & 10 & 14 & 0.672 & 4.24 & 1.141 & 1.768 \\
		$\bm{h_2}$ & 0.707 & 4.24 & 0 & 23 & 1 & 23 & 0.672 & 4.24 & 0.707 & 3.111 \\
		$\bm{h_3}$ & 0.707 & 4.24 & 0 & 23 & 1 & 23 & 0.672 & 4.24 & 0.707 & 3.111 \\
		$\bm{h_4}$ & 0.707 & 4.24 & 1 & 22 & 0 & 24 & 0.74 & 4.17 & 0.707 & 3.111 \\
		$\bm{h_5}$ & 0.707 & 4.24 & 1 & 22 & 0 & 24 & 0.74 & 4.17 & 0.707 & 3.111 \\
		$\bm{h_6}$ & 0.707 & 4.24 & 10 & 13 & 11 & 13 & 0 & 4.911 & 1.485 & 1.697 \\
		$\bm{h_7}$ & 1.414 & 3.536 & 1 & 22 & 1 & 23 & 1.413 & 3.498 & 0 & 3.182 \\
	\bottomrule
	\end{tabular}
	
\end{table*}

In this instance, the $\bm{H}$ matrix is a linear approximation of the Power flow expression. To show that this approximation does not affect the accuracy of the Lemma 1, the case of PMU measurements is considered, namely the systems states are the voltage phasors $V_i$ at each node. The measurements are the equivalent current phasor flows and injections of the previously measured powers. By means of a simple conversion from polar to rectangular coordinates, the complex phasor values can be expressed in terms of their real ($\Re$) and imaginary ($\Im$) parts. In this case, the link between phasor voltages and currents is linear and does not need to be approximated, therefore the $\bm{H}$ matrix is not a linearization but the exact link between states and measurements. If we consider the same line impedances as in figure \ref{fig:3BusSystem}, the resulting $\bm{H}$ matrix when computing the load flow equations can be decoupled in two submatrices:
\begin{equation}
    \bm{H} = 
    \begin{bmatrix}
    \bm{H}_1 & 0\\ 
    0 & \bm{H}_2 
    \end{bmatrix}
\end{equation}

\footnotesize{
\begin{eqnarray}
    &&\bm{H}_1^T = \nonumber \\
    &&\bbordermatrix{
             & V_{1}^{\Im} & V_{2}^{\Im} & V_{3}^{\Im} & I_{12}^{\Re} & I_{13}^{\Re} & I_{31}^{\Re} & I_{32}^{\Re} & I_{23}^{\Re} & I_{1}^{\Re} & I_{3}^{\Re} \cr
            V_{3}^{\Im} & 0 & 0 & 1 & 0 & 1 & -1 & -1 & 1 & 1 & -2 \cr
            V_{2}^{\Im} & 0 & 1 & 0 & 10 & 0 & 0 & 1 & -1 & 10 & 1\cr
            V_{1}^{\Im} & 1 & 0 & 0 & -10 & -1 & 1 & 0 & 0 & -11 & 1\cr
} \nonumber \\
\nonumber \\
&&\bm{H}_2^T = \nonumber \\
&&\bbordermatrix{
             & V_{1}^{\Re} & V_{2}^{\Re} & V_{3}^{\Re} & I_{12}^{\Im} & I_{13}^{\Im} & I_{31}^{\Im} & I_{32}^{\Im} & I_{23}^{\Im} & I_{1}^{\Im} & I_{3}^{\Im} \cr
            V_{3}^{\Re} & 0 & 0 & 1 & 0 & -1 & 1 & 1 & -1 & -1 & 2 \cr
            V_{2}^{\Re} & 0 & 1 & 0 & -10 & 0 & 0 & -1 & 1 & -10 & -1\cr
            V_{1}^{\Re} & 1 & 0 & 0 & 10 & 1 & -1 & 0 & 0 & 11 & -1\cr
} \nonumber
\end{eqnarray}
}

\normalsize{
These submatrices are very similar to the DC approximation $\bm{H}$ matrix, in the sense that all current measurement rows have the same parameters, only the sign changes in one of them which does not affect the leverage point detection method. Furthermore, they are completed by an identity matrix of the voltage measurements. Therefore, the leverage point identification method is applicable to any model type, approximation or real link, as long as the $\bm{H}$ matrix describes a linear system.}

\bibliographystyle{IEEEtran}

\begin{thebibliography}{1}
	    
		\bibitem{Abur-etAl1996} A. Abur and A.G. Exposito, \textit{"Power System State Estimation: Theory and Implementation"}, in IEEE Transactions on Power Systems, Vol. 11, No. 1, February 1996, pp. 216-225.
        
        \bibitem{Mili-etAl1996} L. Mili, M.G. Cheniae, N.S. Vichare and P.J. Rousseeuw, "Robust State Estimation Based on Projection Statistics", in \textit{IEEE Transactions on Power Systems}, Vol. 11, No. 2, May 1996, pp. 1118-1127.
        
		\bibitem{Zhao-etAl2018} J. Zhao and L. Mili, "Vulnerability of the Largest Normalized Residual Statistical Test to Leverage Points," in \textit{IEEE Transactions on Power Systems}, vol. 33, no. 4, pp. 4643-4646, July 2018.
		
        \bibitem{Abur1990} A. Abur, "A Bad Data Identification Method for Linear Programming State Estimation", in \textit{IEEE Transactions on Power Systems}, Vol. 5, No. 3, August 1990, pp. 894-901.
        
        \bibitem{Majumdar-etAl2018} A. Majumdar and B. C. Pal, "Bad Data Detection in the Context of Leverage Point Attacks in Modern Power Networks," in IEEE Transactions on Smart Grid, vol. 9, no. 3, pp. 2042-2054, May 2018.
        
        \bibitem{Abur-etAl2004} A. Abur, A. Gomez-Expsito, \textit{"Power System State Estimation: Theory and Implementation,"} New York Marcel Dekker, 2004.
        
        \bibitem{Paolone-etAl2016} M. Paolone, J.-Y. Le Boudec, S. Sarri and L. Zanni, "Static and recursive PMU-based state estimation processes for transmission and distribution power grids", in \textit{Power System Modelling and Scripting}, edited by F. Milano, Springer Berlin Heidelberg, 2016, Chapter 6.
        
        \bibitem{Ghahremani-etAl2011} E. Ghahremani and I. Kamwa, \textit{"Dynamic State Estimation in Power System by Applying the Extended Kalman Filter With Unknown Inputs to Phasor Measurements,"} in IEEE Transactions on Power Systems, vol. 26, no. 4, pp. 2556-2566, Nov. 2011.
        
        \bibitem{Xu-etAl2013} W. Xu, M. Wang, J. Cai and A. Tang, \textit{"Sparse Error Correction From Nonlinear Measurements With Applications in Bad Data Detection for Power Networks,"} in IEEE Transactions on Signal Processing, vol. 61, no. 24, pp. 6175-6187, Dec.15, 2013.
        
        \bibitem{Abur-etAl1991} A. Abur and M. K. Celik, \textit{"A fast algorithm for the weighted least absolute value state estimation (for power systems),"} in IEEE Transactions on Power Systems, vol. 6, no. 1, pp. 1-8, Feb. 1991.
        
        \bibitem{Gol-etAl2014} M. Göl and A. Abur, \textit{"LAV Based Robust State Estimation for Systems Measured by PMUs,"} in IEEE Transactions on Smart Grid, vol. 5, no. 4, pp. 1808-1814, July 2014.
        
        \bibitem{Rawlings-etAl1998} J. O. Rawlings, S. G. Pantula and D. A. Dickey, "Problem Areas in Least Squares," in \textit{" Applied Regression Analysis: A Research Tool,"}  Springer-Verlag New York, 1998, Chapter 10.
        
        \bibitem{Tan-etAl2014} S. Tan, W. Song, M. Stewart and L. Tong, "LPAttack: Leverage Point Attacks Against State Estimation in Smart Grid", in \textit{IEEE Global Communications Conference}, December 2014, pp. 643-648.
        
        \bibitem{Monticelli2004} A. Monticelli, "State Estimation", in \textit{"Electric Power Systems: A Generalized Approach,"} Kluwer Academic, 2004.
        
        \bibitem{Belsley-etAl1980} D.A. Belsley, E. Kuh,  R.E. Welsch, "Detecting Influential Observations and Outliers" in \textit{"Regression diagnostics"} Wiley, New York, 1980, chapter 2.
        
        \bibitem{Els-etAl1999} S. L. Els, A. D. Els, J. A. Jordaan and R. Zivanovic, "Projection statistics for power system state estimation," in \textit{1999 IEEE Africon. 5th Africon Conference in Africa}, Cape Town, South Africa, 1999, pp. 783-786 vol.2.

        \bibitem{Rousseeuw1990} P.J. Rousseeuw, "Robust Estimation and Identifying Outliers", in \textit{"Handbook of Statistical Methods for Engineers,"} edited by H. Wadsworth Jr, McGraw-Hill ,1990, Chapter 16.
        
        \bibitem{Johnson-etAl2007} R. A. Johnson and D. W. Wichern, "The Multivariate Normal Distribution," in \textit{"Applied Multivariate Statistical Analysis,"} Pearson Prentice Hall, 2007, Chapter 4.
        
        \bibitem{Aggarwal2017} C. C. Aggarwal, "Probabilistic and Statistical Models for Outlier Detection," in \textit{"Outlier Analysis,"} Springer International, 2017, Chapter 2.
        
        \bibitem{Coutto-etAl2014} M. B. Do Coutto Filho, J. C. Stacchini de Souza, M. A. R. Guimaraens, \textit{"Enhanced bad data processing by phasor-aided state estimation,"} in IEEE Trans. Power Syst., vol. 29, no. 5, pp. 2200-2209, Sep 2014.
        
        \bibitem{Asada-etAl2005} E. N. Asada, A. V. Garcia and R. Romero, \textit{"Identifying multiple interacting bad data in power system state estimation,"} IEEE Power Engineering Society General Meeting, 2005, San Francisco, CA, 2005, pp. 571-577 Vol. 1.
        
        \bibitem{Rousseeuw-etAl1991} P.J. Rousseeuw and B.C. Van Zomeren, "Robust Distances: Simulations and Cutoff Values", in Directions in \textit{Robust Statistics and Diagnostics}, Part II, edited by W. Stahel and S. Weisberg, Springer-Verlag, 1991, pp. 195-203.
		
        \bibitem{Mili-etAl1991} L. Mili, V. Phaniraj and P.J. Rousseeuw, "Least Median of Squares Estimation in Power Systems", in \textit{IEEE Transactions on Power Systems}, Vol. 6, No. 2, May 1991, pp. 511-523.
        
		\bibitem{Abur-etAl1997} A. Abur, F. H. Magnago, F. L. Alvarado, "Elimination of leverage measurements via matrix stretching," in \textit{Electrical Power and Energy Systems}, Vol. 19, No. 8, pp. 557-562, 1997.
            
        \bibitem{Mili-etAl1996_tsg} L. Mili, M.G. Cheniae, N.S. Vichare and P.J. Rousseeuw, "Robustification of the Least Absolute Value Estimator by Means of Projection Statistics", in \textit{IEEE Transactions on Power Systems}, Vol. 11, No. 1, February 1996, pp. 216-225.
        
        \bibitem{Cook-etAl1980} R. D. Cook and S. Weisberg, \textit{"Characterizations of an empirical influence function for detecting influential cases in regression,"} in Techno- metrics, 22(4), 495-508, 1980.
	    
	    \bibitem{Barrodale-etAl1973} I. Barrodale and F. D. K. Roberts, "An Improved Algorithm for Discrete l1 Linear Approximation", in \textit{SIAM Journal on Numerical Analysis}, Vol. 10, No. 5, October 1973, pp.. 839-848
        
        \bibitem{Ieee14bus} \textit{"14 Bus Power Flow Test Case,"} in Power Systems Test Case Archive [online] available at: \url{http://labs.ece.uw.edu/pstca/pf14/pg_tca14bus.htm}
        
	    
	    
        

        

        
        
	\end{thebibliography}

\end{document}